\documentclass[twocolumn,prl]{revtex4}
\usepackage{graphicx}
\usepackage{latexsym}

\begin{document}

\title{Ballistic transport in AlAs two-dimensional electrons}
\author{O. Gunawan, Y. P. Shkolnikov, E. P. De Poortere, E. Tutuc, and M.
Shayegan}
\address{Department of Electrical Engineering, Princeton University,
Princeton, New Jersey 08544}
\date{\today}

\begin{abstract}
We report the observation of commensurability oscillations in an
AlAs two-dimensional electron system where two conduction-band
valleys with elliptical in-plane Fermi contours are occupied. The
Fourier power spectrum of the oscillations shows two frequency
components consistent with those expected for the Fermi contours
of the two valleys. From an analysis of the spectra we deduce
$m_l/m_t=5.2\pm0.5$ for the ratio of the longitudinal and
transverse electron effective masses.
\end{abstract}

\pacs{72., 73.23.Ad, 75.47.Jn}

\maketitle

Electrons confined to a modulation-doped AlAs quantum well are an
emerging, high-mobility, two-dimensional electron system (2DES)
with some unique properties that are very different from those of
the more commonly studied GaAs 2DES
\cite{PoortereAPL02,ShkolnikovPRL02,PoorterePRL03}. The AlAs 2D
electrons occupy multiple valleys in the conduction band, each
with a large and anisotropic effective mass, and possess a much
larger effective $g$-factor compared to GaAs 2D electrons. The
combination of these properties has led to the observation of new
phenomena in AlAs 2DES. Examples include magnetic phase
transitions, marked by sharp resistance spikes, at Landau level
crossings in tilted magnetic fields \cite{PoorterePRL03}. Also
observed are fractional quantum Hall states at very high Landau
level filling factors, stabilized possibly because of the
multi-valley occupancy \cite{PoortereAPL02}.  Here we report
measurements of commensurability oscillations (COs) in a
high-mobility AlAs 2DES subjected to a one-dimensional, lateral,
periodic potential modulation.  The 2D electrons in our system
occupy two valleys with elliptical in-plane Fermi contours. The
results demonstrate ballistic transport in a two-valley 2D
system. Through Fourier and partial inverse Fourier analyses of
the oscillations, we disentangle and study the COs of the
electrons in the two valleys, and obtain their amplitude, phase
and scattering time. More importantly, from an analysis of the CO
frequencies, we determine the ratio of the longitudinal and
transverse electron effective masses, a fundamental parameter
that cannot be directly measured from other experiments on our
2DES.

In bulk AlAs, electrons occupy conduction band valleys centered
at the six equivalent X points of the Brillouin zone.  The Fermi
surface consists of six, anisotropic, half-ellipsoids (three
full-ellipsoids); we denote these ellipsoids (valleys) by $X$,
$Y$, and $Z$, according to the direction of their major axes
($x$, $y$, and $z$).  In a narrow (less than $\sim 5$ nm wide)
AlAs quantum well grown on a GaAs (001) substrate, thanks to its
larger mass in the confinement direction, the $Z$ valley
(out-of-plane valley) has the lowest energy and is occupied by
electrons \cite{SiValley}.  For wider well widths, however, the
strain from the lattice mismatch between GaAs and AlAs pushes the
$X$ and $Y$ valleys (the in-plane valleys) down in energy with
respect to the $Z$ valley, so that now the $X$ and $Y$ valleys
are occupied \cite{Smith87,Kesteren89,Maezawa92,Lay93,Yamada94}.
This is the case for the samples we have studied.

\begin{figure}
\includegraphics[scale=.75]{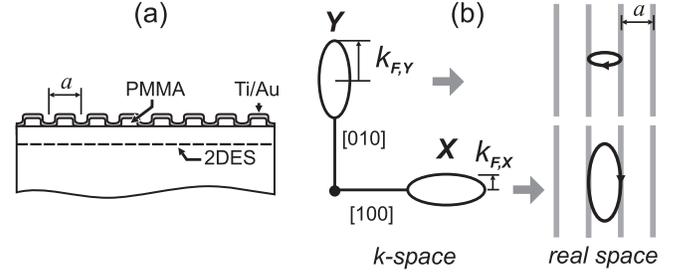} \caption{
(a) Schematic cross section of the device used to measure COs.
Application of a bias to the Ti/Au surface gate with respect to
the 2DES produces a potential modulation which is periodic in the
[100] direction and has period $a$. (b) The AlAs in-plane valleys
$X$ and $Y$ in $k$-space (left), and their corresponding first
resonant CO orbits in real space (right) \cite{COOrbit} are
illustrated. The Fermi wave vectors $k_{F,X}$ and $k_{F,Y}$
relevant for the COs of the $X$ and $Y$ valley are also
indicated.} \label{FigCODevice}
\end{figure}

\label{S_Fig1}

Figure~\ref{FigCODevice} highlights the basic principle of our
study.  Using a grated surface gate, we apply a lateral periodic
potential with period $a$ to the 2DES, and measure the low-field
magneto-resistance ($\rho_{xx}$) along the potential modulation
direction as a function of a perpendicular magnetic field $B$. If
transport is ballistic, $\rho_{xx}$ oscillates with $B$ as the
classical electron cyclotron orbit diameter takes on values that
are multiple integers of $a$
\cite{Weiss89,Winkler89,Gerhardts89,Beenakker89,COOrbit}:
\begin{eqnarray}
  \rho_{xx} \propto \cos \left( 2 \pi f_{CO}/B-\pi/2 \right)
  \label{EqCOrhoxx}
  \\
  f_{CO}=2 \hbar k_F / e a \qquad \qquad  \quad \enskip
  \label{EqfCO}
\end{eqnarray}
where $f_{CO}$ is the oscillation frequency and $k_F$ is the Fermi
wave vector perpendicular to the modulation direction (parallel to
the grating stripes). Note that the oscillations are periodic in
$1/B$. In our AlAs 2DES, there are two in-plane valleys occupied:
$X$ and $Y$. Their cyclotron orbits in real space have the same
shape as their $k$-space orbits but rotated by $90^{\circ}$ as
shown in Fig.~\ref{FigCODevice}(b). If both valleys participate
in the ballistic transport independently, we expect two
superimposed sets of COs whose frequencies are related to the
Fermi wave vectors parallel to the grating stripes as indicated in
Fig.~\ref{FigCODevice}(b):
\begin{equation}
k_{F,X}^{\enskip 2}=2 \pi n_X\sqrt{m_t/m_l} \qquad
k_{F,Y}^{\enskip 2}=2 \pi n_Y\sqrt{m_l/m_t} \label{EqkF}
\end{equation}
where $n_X$ and $n_Y$ are the 2D electron densities for the $X$
and $Y$ valleys respectively. These relations can be combined to
yield:
\begin{equation}
\frac{m_l}{m_t}=\frac{f_{CO,Y}^{\enskip 2}}{f_{CO,X}^{\enskip
2}}\frac{n_X}{n_Y} \label{Eqmlmt}
\end{equation}
implying that, if the valley densities are known, the frequencies
of the COs can be used to directly determine the mass anisotropy
ratio $m_l/m_t$, independent of $a$.

We used samples grown by molecular beam epitaxy on (001) GaAs
substrates \cite{PoortereAPL02}.  The 2DES is confined in an 11
nm-wide AlAs layer located at 110 nm below the surface and
sandwiched between barrier layers of Al$_{0.4}$Ga$_{0.6}$As. It
is modulation doped with a Si delta layer placed at a distance of
75 nm away. A Hall bar mesa was defined on each sample using
standard photolithography and wet etching techniques. The Hall
bar was aligned along the [100] direction so that the major axes
of the two in-plane valleys were either parallel or perpendicular
to the Hall bar. To fabricate the grating patterns, we spun 150
nm of polymethylmetacrylate (PMMA) on top of the sample, and used
electron beam lithography to define an array of PMMA ridges with
periods $a$ equal to 300 and 400 nm. We then deposited 10 nm Ti
and 30 nm Au to form a top gate.  Biasing this top gate with
respect to the 2DES results in a periodic potential modulation in
the 2DES. Using illumination at low temperatures and front/back
gate biasing, we varied the 2DES density between $5$ to
$9\times10^{11}$ cm$^{-2}$, with maximum mobility around 9.3
m$^2$/Vs prior to patterning the grating on top of the sample. We
measured $\rho_{xx}$ in a $^3$He cryostat with a base temperature
of 0.35 K, and used a standard lock-in technique.

\label{S_Fig2}

\begin{figure}
\includegraphics[scale=.72]{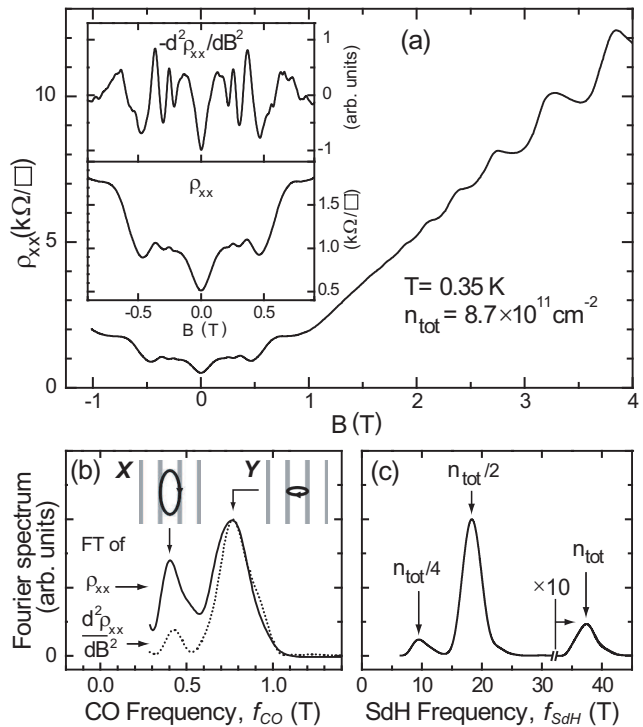} \caption{
(a) Magneto-resistance trace showing COs at low fields ($-1<B<1$
T) and SdHOs at high fields ($B>1.7$ T) for a device with 400 nm
grating period. Insets show the enlarged view of both $\rho_{xx}$
and the numerically determined second derivative
$d^2\rho_{xx}/dB^2$. (b) Fourier power spectra of COs from both
$\rho_{xx}$ (solid curve) and $d^2\rho_{xx}/dB^2$ (dotted curve).
The corresponding first resonant orbits in real space are shown
above the peaks. (c) Fourier power spectrum of SdHOs.}
\label{FigSdHCO}
\end{figure}

A typical $\rho_{xx}$ vs. $B$ trace, taken at $T=0.35$ K and
total density $n_{tot}=8.7\times10^{11}$ cm$^{-2}$, is shown in
Fig.~\ref{FigSdHCO}(a). It exhibits both COs, in the low field
range $-1<B<1$ T, and Shubnikov-de Haas oscillations (SdHOs), at
$B>1.7$ T.  The COs are more clearly seen in the second derivative
($d^2\rho_{xx}/dB^2$) plot shown in the inset of Fig. 2(a).
Fortunately, the COs and SdHOs are well separated in their field
range, thus simplifying their analysis.

The SdHOs provide information regarding the electron densities of
the 2DES and the valleys. In Fig.~\ref{FigSdHCO}(c) we show the
Fourier power spectrum of the SdHOs.  To calculate this spectrum,
we used the $\rho_{xx}$ vs. $1/B$ data for $B>1.7$ T, subtracted a
second-order polynomial background, and multiplied the data by a
Hamming window \cite{DSPHandBook} in order to reduce the
side-lobes in the spectrum. The spectrum exhibits three clear
peaks, marked in Fig. 2(c) as $n_{tot}$, $n_{tot}/2$ and
$n_{tot}/4$.  The SdHO frequencies multiplied by $e/h$ give the
2D density ($e$ is the electron charge and $h$ is the Planck's
constant). We associate the $n_{tot}$ peak with the total
density, as the position of this peak multiplied by $e/h$ indeed
gives the total 2DES density which we independently determine
from the Hall coefficient. For the data shown in Fig. 2, we
deduce $n_{tot} = n_X + n_Y = 8.7\times10^{11}$cm$^{-2}$. The
presence of the $n_{tot}/4$ peak indicates the spin and valley
degeneracy of the system \cite{SdHPeaks}.

Figure ~\ref{FigSdHCO} (b) shows the Fourier power spectra of COs
calculated using $\rho_{xx}$ and $d^2\rho_{xx}/dB^2$ vs. $1/B$
data in the $0.1<B<1$ T range. Both spectra exhibit two clear
peaks at $f_{CO,X}$ and $f_{CO,Y}$, which we associate with the CO
frequencies of the $X$ and $Y$ valleys, respectively. If we
assume that the two valleys have equal densities, we can use
Eq.~(\ref{Eqmlmt}) to immediately find  $m_l/m_t\simeq4.4$. This
value, however, is inaccurate because there is a small but finite
imbalance between the $X$ and $Y$ valley densities in our sample.
Such imbalances can occur because of anisotropic strain in the
plane of the sample and are often present in AlAs 2DESs. Note
that the Fourier spectrum of the SdHOs cannot resolve small valley
density imbalances. As detailed in the next paragraph, we analyze
the dependence of CO frequencies on density to deduce the
imbalance between the valley densities, and also to determine the
$m_l/m_t$ ratio more accurately.

\label{S_Fig3}

\begin{figure}
\includegraphics[scale=.85]{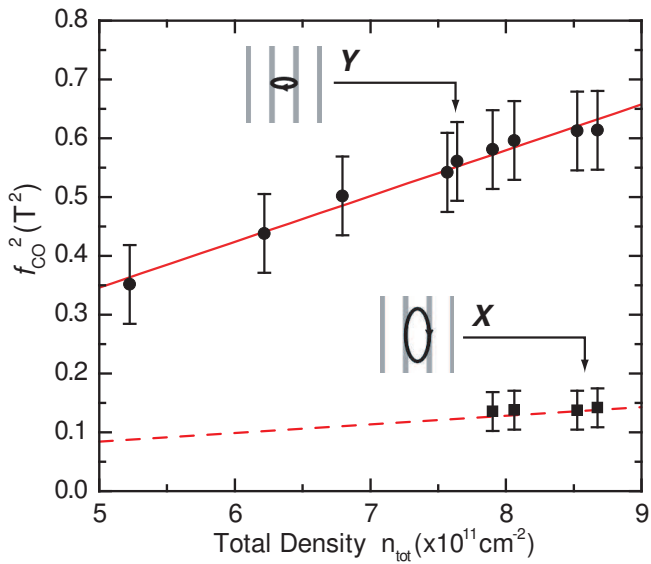} \caption{
Density dependence of the CO frequencies for the $Y$ (circles)
and $X$ valleys (squares). The line through the circles is a
least-squares fit to the data; its slope determines the ratio
$m_l/m_t$ and its intercept the density difference $\Delta n$ of
the two valleys. The dashed line is described in the text.}
\label{FigfCO2}
\end{figure}

Figure~\ref{FigfCO2} summarizes the density dependence of our CO
frequencies. Denoting the difference between the valley densities
by $\Delta n=n_Y-n_X$, we can rewrite Eqs. (\ref{EqfCO}) and
(\ref{EqkF}) as:

\begin{eqnarray}
f_{CO,Y}^{\enskip 2}=\frac{h^2}{\pi e^2 a^2} \sqrt{m_l\over m_t}
\left(n_{tot}+\Delta n\right) \label{EqfCOYvsN}
\\
f_{CO,X}^{\enskip 2}=\frac{h^2}{\pi e^2 a^2} \sqrt{m_t\over m_l}
\left(n_{tot}-\Delta n\right). \label{EqfCOXvsN}
\end{eqnarray}
From the slope and intercept of the $f_{CO}^{\enskip 2}$ vs.
$n_{tot}$ plots, we can deduce the $m_l/m_t$ ratio and $\Delta
n$. Concentrating on the $Y$ valley, a least-squares fit of
$f_{CO,Y}^{\enskip 2}$ data points (circles in
Fig.~\ref{FigfCO2}) to a line leads to values $m_l/m_t=5.2\pm0.5$
and $\Delta n=(-0.6\pm0.4)\times 10^{11}$ cm$^{-2}$. Note that
such a small value of $\Delta n$ is consistent with the nearly
valley-degenerate picture deduced from the existence of the
$n_{tot}/4$ peak in the SdH frequency spectrum
[Fig.~\ref{FigSdHCO}(c)].

The above determination of the $m_l/m_t$ ratio is based on the
density dependence of $f_{CO,Y}$ only and does not use the
measured $f_{CO,X}$. As a consistency check, we can use
Eq.~(\ref{EqfCOXvsN}) to predict $f_{CO}$ for the $X$ valley using
$m_l/m_t$ and $\Delta n$ deduced from the above analysis of
$f_{CO,Y}^{\enskip 2}$. This prediction, shown as a dashed line in
Fig.~\ref{FigfCO2}, is in good agreement with the measured
$f_{CO,X}^{\enskip 2}$ (solid squares). This consistency confirms
that we are indeed observing COs for both valleys.

We have repeated similar experiments in a sample from a different
wafer, containing a 2DES confined to a 15 nm wide AlAs quantum
well. In this sample only the COs of the $Y$ valley could be
reliably determined. By performing similar analysis using
Eq.~(\ref{EqfCOYvsN}), in the density range from $5.5$ to
$9\times10^{11}$ cm$^{-2}$, we deduce $m_l/m_t \simeq 5.4$, in
good agreement with the results presented here.

At this point it is worthwhile emphasizing that the COs described
here uniquely probe the $m_l/m_t$ ratio \cite{FaradayRot}.
Conventional experiments that probe the effective mass, such as
cyclotron resonance or measurements of the temperature dependence
of the amplitude of the SdHOs, lead to a determination of the
cyclotron effective mass, $m_{CR}$.  In a 2DES with an elliptical
Fermi contour, $m_{CR}$ is equal to $(m_l m_t)^{1/2}$, and
therefore provides information complimentary to the $m_l/m_t$
ratio, so that $m_l$ and $m_t$ can be determined. In fact, using
the measured $m_{CR}=0.46 m_e$ in AlAs 2DESs \cite{CyclotronRes},
we can use the $m_l/m_t=5.2\pm0.5$ ratio to deduce $m_l =
(1.1\pm0.1)m_e$ and $m_t = (0.20\pm0.02) m_e$. These values are in
good agreement with the (theoretical) value of $m_t=0.19m_e$ that
is calculated in Ref.~\cite{FaradayRot}, and $m_l=1.1m_e$ that is
deduced from the Faraday rotation measurements \cite{FaradayRot};
they also agree well with the results of the majority of
theoretical and experimental determinations of the effective mass
in AlAs.

\label{S_Fig4}
\begin{figure}
\includegraphics[scale=0.8]{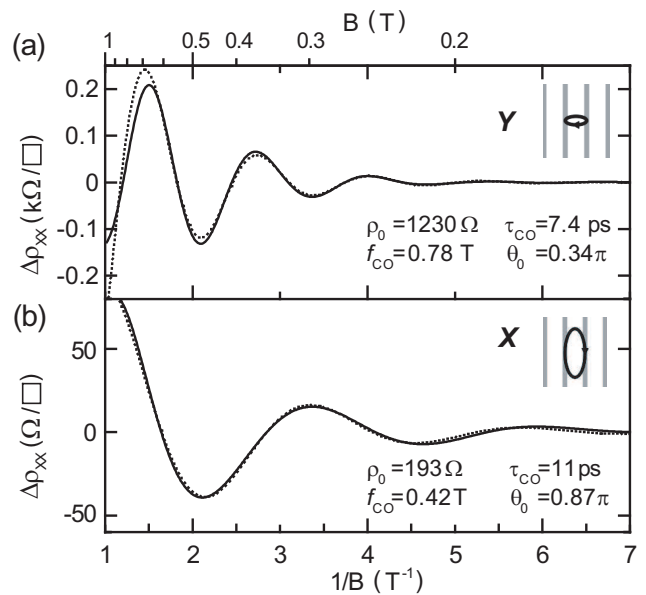} \caption{\label{FigCODecompo}
Results of the inverse Fourier decomposition of the COs of
Fig.~\ref{FigSdHCO} for the $Y$ and $X$ valleys. The dotted
curves show the best fits of Eq. (\ref{EqCODecay}) using the
indicated parameters.}
\end{figure}

We  proceed to extract more information, such as the amplitude,
phase, and scattering time for the COs of each valley by
performing partial inverse Fourier analysis.
Figure~\ref{FigCODecompo} summarizes the results of such
analysis. The Fourier power spectrum shown in Fig.
\ref{FigSdHCO}(b) is separated into two regions chosen to isolate
the two CO peaks. The region corresponding to COs of $Y$ valley
($0.57<f_{CO}<1.21$ T) is inverse Fourier transformed and divided
by the original window function. The result is shown as the solid
curve in Fig.~\ref{FigCODecompo}(a). The region corresponding to
the COs of $X$ valley ($0.29<f_{CO}<0.57$ T), is analyzed in the
same manner and the result is shown by the solid curve in Fig.
\ref{FigCODecompo}(b) \cite{COBackground}.

We fit the deduced COs for each valley to a simple expression
that assumes the amplitude of the COs decreases exponentially with
$1/B$:
\begin{equation}
\Delta \rho_{xx} \propto \rho_0 \exp \left(-\pi/\omega_c\tau_{CO}
\right) \cos\left(2 \pi f_{CO}/B-\theta_0\right) \label{EqCODecay}
\end{equation}
where $\rho_0,\tau_{CO}, f_{CO},$ and $\theta_{0}$ are the
fitting parameters; $\omega_c=eB/m_{CR}$ is cyclotron frequency
with $m_{CR}=(m_l m_t)^{1/2} =0.46m_e$. The exponential term of
expression (\ref{EqCODecay}) is analogous to the Dingle factor
used to describe the damping of the SdHOs' amplitude with
increasing $1/B$, and has been used successfully to fit COs in
GaAs 2D electrons \cite{LuPRB98} and holes \cite{LuPRB99}. In
Fig.~\ref{FigCODecompo} the results of the best fits are shown as
dotted curves along with their fitting parameters. The best-fit
$\theta_0$ for the COs of $Y$ and $X$ valleys are $0.34\pi$ and
$0.87\pi$ respectively, in reasonable agreement with the expected
value of 0.5$\pi$  (the relative phase errors are 8\% and 18\% of
$2\pi$). This consistency affirms that the reconstructed
oscillations faithfully represent the COs of the two in-plane
valleys. The amplitude of the oscillations for the $Y$ valley is
significantly larger than for the $X$ valley as expected from the
shorter real-space, resonant orbital trajectories for this valley
[Fig.~\ref{FigCODevice}(b)]. On the other hand, the scattering
times, $\tau_{CO}$, that we deduce from the fits are comparable
for the two valleys, suggesting that scattering is nearly
isotropic.

We also deduce two other scattering times: the quantum lifetime
$\tau_{SdH}$ and the mobility scattering time $\tau_{\mu}$, and
compare them with $\tau_{CO}$. From fitting the $B$ dependence of
the amplitude of the SdHOs to the damping factor $\exp
\left(-\pi/\omega_c\tau \right)$, we obtain $\tau_{SdH}=0.76$ ps.
The mobility scattering time is $\tau_{\mu}=24$ ps, determined
from the mobility of the same sample prior to patterning. Similar
to CO experiments in other 2D carrier systems
\cite{LuPRB98,LuPRB99}, we observe $\tau_{SdH} <\tau_{CO} <
\tau_{\mu}$. This observation can be qualitatively understood
considering the sensitivity of these $\tau$ to the scattering
angle \cite{LuPRB98}: $\tau_{\mu}$ is the longest since the
mobility is least sensitive to small-angle scattering, while
$\tau_{SdH}$ is the shortest because the SdHOs are sensitive to
all scattering events.

\label{S_conclusion} We close by reflecting on possible future
studies where ballistic transport in AlAs 2DESs can be applied.
One possibility is to apply the periodic potential along
different in-plane directions and map out the Fermi contour shape
\cite{Heremans94}. A second area involves studies of ballistic
\textit{spin polarized } currents. Because of the large values of
effective mass and $g$-factor \cite{PapadakisPRB99} of AlAs 2D
electrons, it is sufficient to apply an in-plane magnetic field
of only a few T to fully spin-polarize the electrons. Such
studies are of substantial current interest as they have
relevance for spintronic devices. A third area concerns chaotic
transport in anti-dots; the highly anisotropic electron Fermi
contours of AlAs 2D electrons should lead to interesting
phenomena, examples of which have been reported recently in GaAs
2D hole systems which also possess non-circular Fermi contours
\cite{Zitzlsperger03}.

We thank the NSF and ARO for supporting this work, and J.\ J.\
Heremans and R. Winkler for illuminating discussions.


\begin{references}
\bibitem{SiValley}
This is similar to (001) Si-MOSFETs, except that in the Si case
the 2D electrons occupy $two$ out-of-plane valleys at equivalent
points along the Brillouin zone $\Delta$ lines.
\bibitem{COOrbit}
More precisely, as Eq.~(\ref{EqCOrhoxx}) indicates, a maximum in
$\rho_{xx}$ is seen whenever the cyclotron orbit diameter equals
$a(p+1/4)$, where $p$ is an integer.  In
Fig.~\ref{FigCODevice}(b), for simplicity, we schematically show
the condition for the first resonance as when the diameter is
equal to $a$.
\bibitem{SdHPeaks}
As detailed in Ref.~\cite{ShkolnikovPRL02}, the spin and valley
degeneracies are lifted at higher $B$, leading to the presence of
the $n_{tot}/2$ and $n_{tot}$ peaks in the Fourier spectrum.
\bibitem{COBackground}
If we subtract the sum of the CO traces deduced for the individual
valleys (shown in Fig.~\ref{FigCODecompo}) from the orginal CO
trace of Fig.~\ref{FigSdHCO}(a), we obtain a featureless,
non-oscillatory background signal.
\end{references}
\end{document}